\documentclass[useAMS,usenatbib]{mn2e}
\usepackage{graphicx}
\usepackage{amsmath,amssymb}
\title[Competitive accretion in the protocluster G10.6-0.4?]{Competitive accretion in the protocluster G10.6-0.4?}
\author[Tie Liu, Yuefang Wu, Jingwen Wu, Sheng-Li Qin, Huawei Zhang]{Tie Liu$^{1}$\thanks{E-mail:
liutiepku@gmail.com}, Yuefang Wu$^{1}$\thanks{E-mail:
yfwu.pku@gmail.com}, Jingwen Wu$^{2}$, Sheng-Li Qin$^{3}$, Huawei Zhang$^{1}$.\\
$^{1}$Department of Astronomy, Peking University, 100871, Beijing China\\
$^{2}$Jet Propulsion Laboratory, 4800 Oak Grove Dr., Pasadena, CA 91101, U.S.A\\
$^{3}$I. Physikalisches Institut, Universit\"at zu K\"oln, Z\"ulpicher Str. 77, 50937 K\"oln, Germany }

\begin{document}


\pagerange{\pageref{firstpage}--\pageref{lastpage}} \pubyear{2012}

\maketitle

\label{firstpage}

\begin{abstract}
We present the results of high spatial resolution observations at 1.1 mm waveband, with the Submillimetre Array (SMA), towards the protocluster G10.6-0.4. The 1.1 mm continuum emission reveals seven dense cores, towards which infall motions are all detected with the red-shifted absorption dips in HCN (3--2) line. This is the first time that infall is seen towards multiple sources in a protocluster. We also identified four infrared point sources in this region, which are most likely Class 0/I protostars. Two jet-like structures are also identified from Spitzer/IRAC image. The dense core located in the centre has much larger mass than the off-centre cores. The clump is in overall collapse and the infall motion is supersonic. The standard deviation of core velocities and the velocity differences between the cores and the cloud/clump are all larger than the thermal velocity dispersion. The picture of G10.6-0.4 seems to favor the "competitive accretion" model but needs to be tested by further observations.
\end{abstract}


\section{Introduction}
There are two most promising models to account for high-mass star formation (M $>$8 M$_{\sun}$). One is called the "monolithic collapse and disk accretion" \citep{york02,mck03}, and the other is "competitive accretion" \citep{bonn02,bonn04,bonn06}. The former one suggests that high-mass stars form directly from isolated massive gas clumps like low-mass star formation but with a much larger accretion rate. The latter claims that high-mass stars form in the centre of a cluster through competitive gathering gas mass with the other off-centre protostars.

The competitive accretion model was proposed based on the observational fact that most stars are formed in clusters, and most massive stars reside in the
centre, i.e, mass segregation \citep{bonn02}. In this model, the mutual potential of the protostars in the cluster funnels gas down to the centre, which results in higher gas densities and higher accretion rates for the few protostars located there. Thus the protostars located in the centre can accrete more and more gas and form massive stars. However, till now, no convincing observational evidences support the "competitive accretion" model. What is the mass segregation in protoclusters? How do the protostars interact with each other? Do the protostars competitively accumulate masses under a common potential? To distinguish these two models, high spatial resolution observations towards protoclusters are critically needed.

Located at a distance of 6 kpc, the dense clump G10.6-0.4 has been widely studied from centimeter to sub-millimetre wavelengths at various spatial resolutions (e.g., \citet{liu11} and the references therein). The previous centimeter observations have pointed out burst of massive star formation in the central UC~H{\sc ii} region \citep{ho86,keto87,keto88,soll05}. A flatten rotating torus is found in the central parsec region of the dense clump \citep{hoha86,liu10a}. Gas infalling was also revealed in the form of redshifted absorption seen against the continuum source \citep{keto87,keto88,soll05}. Multiple outflows were detected in the $^{12}$CO (2--1) and HCN (3--2) lines \citep{liu10b}. All the previous observations indicate a high-mass star cluster is forming in a dense clump \citep{liu11}. Thus it is a good target for testing different models for high mass star foramtion with sufficiently high spatial resolution observations and sensitive molecular tracers for infall motion. In previous VLA observations \citep{keto87,keto88,soll05}, the NH$_{3}$ lines used as a tracer for infall motions only revealed the gas flow in the inner warm part near the central ultracompact H{\sc ii} region, which cannot provide information of the infall motions in the other less dense and colder parts of the dense clump. In this paper, with the HCN (3--2) line, which is a good tracer for infall motions \citep{liu11a,liu11b,wu03}, we can reveal the pattern of infall motions in the whole dense clump and uncover the accretion mode for the formation of a high-mass star cluster.

\section{Observations}

The observations of G10.6+0.4 were carried out with the SMA in
July 2005 with seven antennas in its compact configuration and in
September 2005 with six antennas in its extended configuration. The integration time on source before flagging for the compact array and extended array is 139 minutes and 108 minutes, respectively. The phase reference centre of both observations was R.A.(J2000)~=~18$^{\rm h}$10$^{\rm m}$28.698$^{\rm s}$ and
DEC.(J2000)~=~-$19\degr55\arcmin48.68\arcsec$. The 345 GHz receivers were tuned to 265 GHz for the lower sideband (LSB)
and 275~GHz for the upper sideband (USB). The frequency resolution
across the spectral band is 0.8125~MHz or $\sim$1 km s$^{-1}$ for
both configurations.

In both configuration observations, Jupiter, Uranus
and QSO 3c454.3 were observed for antenna-based bandpass correction.
QSOs 1741-038 and 1908-201 were employed for antenna-based gain correction.
Uranus was observed for flux-density calibration.

Miriad was employed for calibration and imaging \citep{sau95}. The 1.1 mm continuum data were acquired by
averaging all the line-free channels over both the 2 GHz of upper
and lower spectral bands. MIRIAD task "selfcal" was employed to
perform self-calibration on the continuum data.
The gain solutions from the self-calibration were applied to the
line data.

The synthesized beam size and 1 $ \sigma$ rms of the dust emission observed in the compact configuration
is $3\arcsec.47\times2\arcsec.89$ (PA=-45.7$\degr$) and 30 mJy~beam$^{-1}$, respectively. The continuum data combined from both configurations yield a
synthesized beam of 1$\arcsec.24\times1\arcsec.11$, P.A.=-61$\degr$.6, and 1 $ \sigma$ rms of 10 mJy~beam$^{-1}$ in the uniform weighted map.

The HCN (3--2) (265.886 GHz) and CH$_{3}$OH (5$_{2,3}$-4$_{1,3}$) (266.838) lines were observed at the lower band, while the CH$_{3}$CN (15-14) transitions were
observed at upper band. The typical 1 $ \sigma$ rms of line emission is $\sim$150 mJy~beam$^{-1}$.

IRAC data were also retrieved from the database of GLIMPSE.

\section{Results}

\subsection{1.1 mm continuum emission}
The 1.1 mm continuum image obtained from the SMA compact array is shown in red solid contours
in the upper panel of Figure 1. As well as the central elongated clump, two small clumps were found in the north-east and north-west. The two UC~H{\sc ii} regions identified by \citet{liu10b} were denoted as black plus symbols. The emission peak of the 1.1 mm continuum emission coincides with the central bright UC~H{\sc ii} region. The deconvolved size of the large clump is 3$\arcsec.67\times1\arcsec.07$ (P.A.=-64.1$\degr$). By combining the data of the compact and extended configurations of SMA, we obtained a much higher spatial resolution (1$\arcsec.24\times1\arcsec.11$, P.A.=-61$\degr$.6) image of the 1.1 mm dust continuum emission. As shown in the lower panel of Figure 1, The large clump detected by the compact array is further resolved into seven dense cores, which are denoted from "A" to "G". We applied two dimensional gaussian fits towards these cores. The positions, integrated intensity and radii of these cores are listed in columns 2-4 of Table 1.

\begin{figure}
\begin{center}
\begin{tabular}{c}
\includegraphics[scale=0.45]{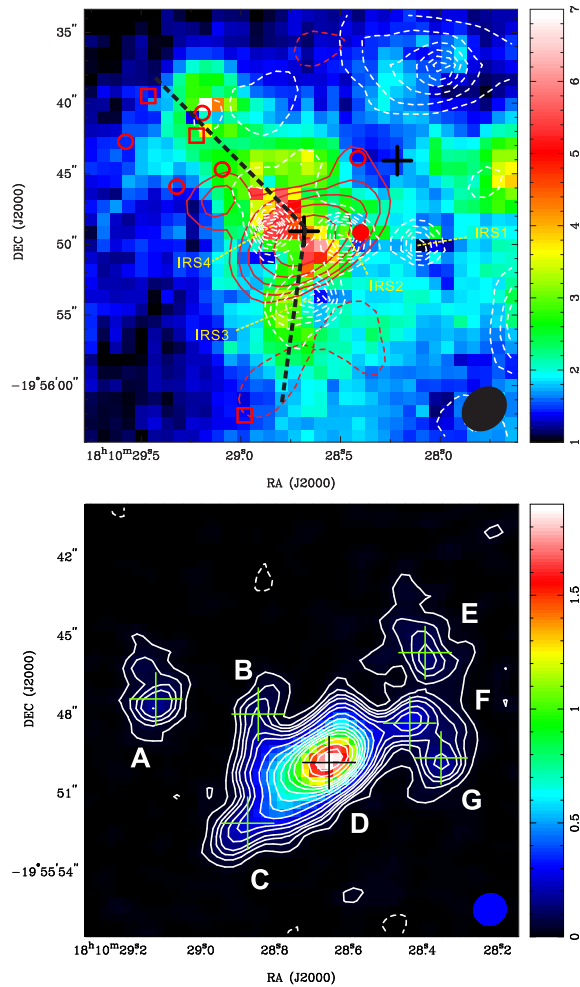}
\end{tabular}
\end{center}
\caption {Upper: The flux ratio map of [4.5]/[3.6] from Spitzer/IRAC is shown in color scale. The white dashed contours represent the distribution of 8 $\micron$ emission. The contour levels are from 400 to 1200 mJy/sr in a step of 100 mJy/sr. The 1.1 mm continuum from the compact configuration is shown as red solid contours. The contour levels are (-5,5,10,20,40,80,160)$\times$0.03 Jy/beam (1 $\sigma$). The two black plus symbols mark the positions of the UC~H{\sc ii} region. The red open circles and open boxes mark the locations of the blueshifted and redshifted $^{12}$CO outflow interaction signatures \citep{liu10b}. The red solid circle is the location of the HCN (3--2) outflow. The names of the four infrared point sources are labeled as IRS1-4. The two dashed long lines denote the directions of elongated structures with [4.5]/[3.6]$>$2. Lower: The 1.1 mm continuum from combining the compact and extended configurations is shown in white solid contours and in color scale. The contour levels are (-3,3,6,9,12,18,24,36,48,60,72,84,96,120,144,168)$\times$0.01 Jy beam$^{-1}$ (1 $\sigma$). The positions of the cores are marked with "crosses" and labeled from "A" to "G".}
\end{figure}

Assuming that the dust emission is optically thin and the dust temperature equals the rotational temperature (87 K) derived from CH$_{3}^{13}$CN (12-11) \citep{belt11},
the masses of seven cores can be obtained with the formula
$M=S_{\nu}D^{2}/\kappa_{\nu}B_{\nu}(T_{d})$, where $S_{\nu}$ is the
flux of the dust emission at 1.1 mm, D is the distance, and $B_{\nu}(T_d)$ is the Planck
function. Here the ratio of gas to dust is taken as 100. The dust opacity $\kappa_{\nu}=\kappa_{1300}(\frac{1300
\micron}{\lambda})^{\beta}$=0.011 cm$^{2}$g$^{-1}$, where $\beta$=1.5 is the dust opacity index
and $\kappa_{1300}=0.009$ cm$^{2}$g$^{-1}$ is the dust opacity at
1300 $\micron$ derived from a gas/dust model with thin ice mantles and a gas density of $\sim10^{6}$ cm$^{-3}$
\citep{oss94}. The dust opacity depends on the size of the dust grain and has great effect on the mass measurement. If we adopt the dust opacity index of 2, the inferred mass should be reduced by $\sim$10\%. Since the central core (D) is a UC~H{\sc ii} region, which should have a higher temperature than the clump averaged temperature derived from CH$_{3}^{13}$CN (12-11) lines, we derived the mass and volume density for it by assuming a dust temperature of 161 K (see section 4.1 for details). The inferred masses and surface density of these cores are listed in columns 5 and 6 of Table 1. The masses of the cores range from 19 to 207 M$_{\sun}$.

\begin{table*}
 \centering
 \begin{minipage}{140mm}
  \caption{Parameters of each dense core.}
  \begin{tabular}{@{}ccccccccccc@{}}
  \hline
 Name & Offset &S$_{\nu}$  & R  & M& N   &V$_{CH_{3}OH}$  &$\sigma_{NT}$   \\
 & ($\arcsec,\arcsec$)  &
(Jy)  & ($10^{3}$ AU)  & (M$_{\sun}$) & (10$^{24}$ cm$^{-2}$) & (km~s$^{-1}$)    & (km~s$^{-1}$) \\
 \hline
A & (6.0,1.3)      & 0.37(0.06)    &4.3(0.5)    & 31(5)  & 1.0(0.4)     & -7.50(0.06) & 1.72(0.06) \\
B & (2.1,0.7)      & 0.23(0.03)    &2.7(0.3)    & 19(2)  & 1.6(0.5)     & -3.73(0.23) & 1.80(0.20) \\
C & (2.5,-3.5)     & 0.56(0.03)    &3.5(0.2)    & 46(2)  & 2.3(0.4)     & -5.10(0.07) & 1.43(0.07) \\
D & (-0.3,-1.2)    & 4.81(0.31)%
  \footnote{Since the integrated flux density at 1.3 cm is as high as 2.5 Jy \citep{soll05}, the free-free contribution
to the 1.1 mm can't be ignored. Assuming optically thin free-free emission ($S_{\nu}\propto\nu^{-0.1}$) \citep{belt11}, 1.99 Jy of free-free emission was subtracted from the total flux 6.8 Jy.}    & 6.0(0.3)  &207(17)\footnote{This value is calculated by assuming that the dust temperature is 161 K.} & 3.7(0.3)     & -2.74(0.05) & 2.17(0.04)   \\
E & (-4.3,3.0)     & 0.50(0.07)    &6.4(0.8)    & 41(6)  &0.6(0.2)     & -0.07(0.14) & 3.13(0.13)  \\
F & (-3.7,0.4)     & 0.60(0.07)    &3.7(0.4)    & 50(6)  &2.3(0.8)     & -1.07(0.11) & 2.29(0.11) \\
G & (-4.9,-1.0)    & 0.37(0.04)    &3.5(0.3)    & 31(3)  &1.7(0.5)     & -2.35(0.15) & 1.92(0.13) \\
\hline
\end{tabular}
\end{minipage}
\end{table*}

\subsection{Infrared emission}
Four infrared point sources locating at the central part are revealed by the 8 $\micron$ emission presented as white dashed contours in the upper panel of Figure 1, which are denoted as IRS1-4. IRS1 locates west and is not associated with the 1.1 mm dust emission. The other three infrared point sources form a triangular system, whose geometric centre coincides with the 1.1 mm emission peak. All the four infrared point sources have colors [3.6]-[4.5]$\geq$0.4 and [5.8]-[8.0]$\geq$1.1,
indicating they are most likely protostars (Class 0/I objects) \citep{qiu08}.

The flux ratio map of [4.5] to [3.6] $\micron$ emission is shown in color scale in the upper panel of Figure 1. Two elongated structures with [4.5]/[3.6]$>$2 are revealed and marked by the two long dashed lines. Despite of the elongated structures, two red spots with ratios larger than 4 are found near the 1.1 mm emission peak. The flux ratio of [4.5] to [3.6] in the jets is comparable or higher than $\sim$1.5 in contrast to the stars \citep{ta10}. The two elongated structures may depict the jets generated from the protocluster G10.6-0.4. The roots of the two elongated structures connect with two red spots, indicating the interaction between the shocked jets and the envelope. We also noticed that the $^{12}$CO outflow knots distribute along the elongated structures and mainly locate at the outer layer or the tip of the jets, indicating the $^{12}$CO outflow gas is entrained and accelerated by the collimated jets.

\subsection{Line emission}
The source averaged HCN (3--2) and CH$_{3}$OH (5$_{2,3}$-4$_{1,3}$) lines are shown in Figure 2. As marked by the red dashed lines, red-shifted absorption dips were detected in the HCN (3--2) lines, indicating gas infall is taking place in this region. This is the first time
that infall is seen towards multiple sources in a protocluster. Only core "F" and core "G" show strong high velocity blue wings. The CH$_{3}$OH (5$_{2,3}$-4$_{1,3}$) line is single-peaked towards all the cores. The systemic velocities of the CH$_{3}$OH (5$_{2,3}$-4$_{1,3}$) line were obtained from gaussian fit and listed in column 7 of Table 1. Roughly speaking, we found the CH$_{3}$OH (5$_{2,3}$-4$_{1,3}$) lines are red-shifted towards the western cores ("E", "F", and "G") and blue-shifted towards the eastern cores ("A", "B", and "C").

\begin{figure}
\begin{center}
\begin{tabular}{c}
\includegraphics[scale=0.4]{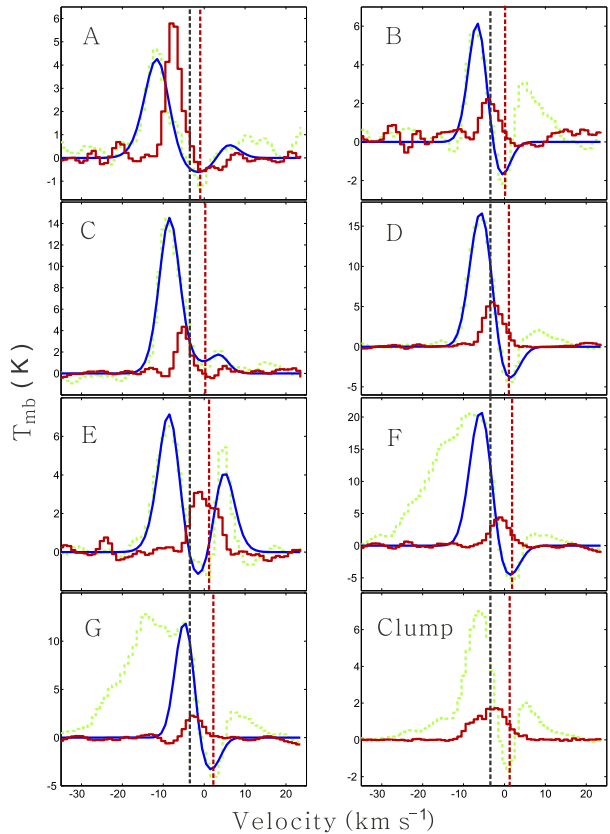}
\end{tabular}
\end{center}
\caption{Spectra averaged over each core. green: HCN (3--2), red: CH$_{3}$OH (5$_{2,3}$--4$_{1,3}$). The last panel presents the clump averaged spectra obtained only from compact configuration. The black dashed lines mark the position (-3 km s$^{-1}$) of the systemic velocity. The red dashed lines mark the absorption dips. The best "two-layer" model fits towards the dense cores are exhibited as blue solid lines.}
\end{figure}

\begin{table*}
 \centering
 \begin{minipage}{140mm}
  \caption{Best fitting results for the HCN (3--2) line with the enhanced two layer models}
  \begin{tabular}{@{}cccccccccccccccccc@{}}
  \hline
 Name & $\tau_{0}$ &f  & J$_{c}$ & J$_{f}$ & J$_{r}$ & V$_{cont}$  &$\sigma_{HCN}$ &V$_{rel}$ \\
 &   &
  & ( K ) & ( K ) & ( K ) & (km~s$^{-1}$)    & (km~s$^{-1}$)  & (km~s$^{-1}$)  \\
 \hline
A & 3.0(0.2)     & 0.4(0.1)    &44.3(1.7)  &18.1(0.8)  & 19.1(0.9)   & -3.0(0.3) & 4.9(0.1) & 1.3(0.1)  \\
B & 0.5(0.1)     & 0.4(0.1)    &49.3(1.9)  &8.1(0.9)   & 38.7(1.9)   & -3.9(0.1) & 2.5(0.1) & 1.5(0.1)  \\
C & 2.8(0.1)     & 0.5(0.1)    &54.8(1.8)  &25.2(0.8)  & 51.0(2.0)   & -2.3(0.1) & 3.6(0.1) & 1.6(0.1)    \\
D & 0.5(0.1)     & 0.4(0.1)    &91.3(1.6)  &22.8(0.6)  & 70.8(0.9)   & -2.5(0.1) & 2.7(0.1) & 2.8(0.1)   \\
E & 2.0(0.1)     & 0.5(0.2)    &43.1(17.8)  &15.3(8.7)  & 64.5(20.0)  & -1.9(0.1) & 4.2(0.1) & 0.3(0.1)  \\
F & 0.7(0.1)     & 0.3(0.1)    &65.9(2.3)  &8.1(0.7)   & 63.6(1.1)    & -2.5(0.1) & 2.7(0.1) & 2.7(0.2)   \\
G & 1.1(0.1)     & 0.4(0.1)    &44.6(1.3)  &11.4(0.5)  & 32.0(0.9)    & -1.6(0.1) & 2.3(0.1) & 2.5(0.1)  \\
\hline
\end{tabular}
\end{minipage}
\end{table*}

\section{Discussion}

\subsection{Fragmentation of the clump}
As shown in Figure 1, the dust clump fragments into seven dense cores. The central core has the largest mass, while the off-centre cores have much smaller masses. Figure 3 presents the integrated intensity map of HCN (3--2). One can find that the emission component of HCN (3--2) also fragments into several gas cores. The absorption of HCN (3--2) is mainly associated with the central dust core. The outflow emission of HCN (3--2), as shown in blue contours, reveals a compact lobe, which is located at the west of the central dust core and is only associated with dust core "F" and "G", indicating that the previously discovered HCN outflow \citep{liu10b} in this region is driven by the forming protostars in core "F" and core "G". In Figure 4, we presents the Position-Velocity (P-V) cut of HCN (3--2) along the west-east direction through the central dust core. One can see the HCN (3--2) components of outflow emission, core emission and absorption are well separated in velocity space. As marked by the green arrows in Figure 4, the gas emission traced by HCN (3--2) fragments into individual cores with a velocity gradient from west to east. The outflow component also shows a chain of knots from -30 to -10 km~s$^{-1}$. From the P-V cut, one can also see that the HCN (3--2) outflow is only located to the west of the central core.

\begin{figure}
\begin{center}
\begin{tabular}{c}
\includegraphics[scale=0.28,angle=90]{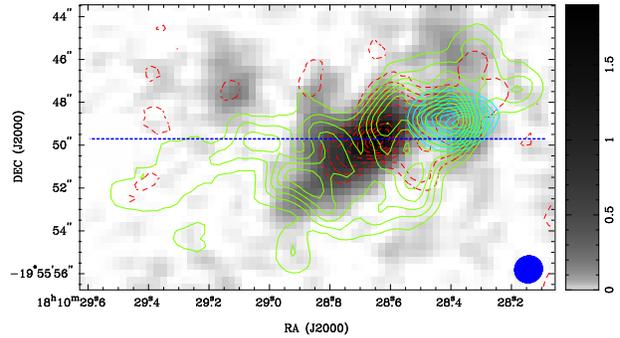}
\end{tabular}
\end{center}
\caption{Blue solid contours: outflow emission of HCN (3--2) integrated from -30 to -15 km~s$^{-1}$. The contours are from 10\% to 90\% of the peak emission. Green solid contours: clump emission of HCN (3--2) integrated from -10 to 3 km~s$^{-1}$. The contours are from 10\% to 90\% of the peak emission. Red dashed contours: absorption of HCN (3--2) integrated from -1 to 5 km~s$^{-1}$. The contours are from -9 to -1 Jy~beam$^{-1}$~km~s$^{-1}$. The background image is the 1.1 mm continuum emission in log scale. The horizontal dashed line is the orientation of the PV cut in Figure 4. }
\end{figure}

\begin{figure}
\begin{center}
\begin{tabular}{c}
\includegraphics[scale=0.3,angle=90]{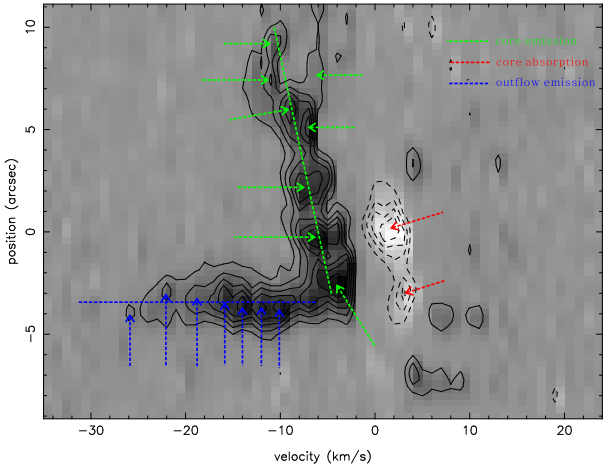}
\end{tabular}
\end{center}
\caption{Position-velocity diagram of HCN (3--2) along the W-E direction as depicted by the long dashed line in Figure 3. The contours are from -60\% to 90\% in steps of 15\% of the peak emission.}
\end{figure}

In Figure 5, we present the Moment 0 contours and Moment 1 images of CH$_{3}$OH (5$_{2,3}$--4$_{1,3}$). The red contours in the upper panel shows the integrated intensity map of CH$_{3}$OH (5$_{2,3}$--4$_{1,3}$) emission obtained from the compact array. As depicted by the dashed lines, an "X-shape" structure consisted of five gas cores is revealed in the integrated intensity map. A velocity gradient is seen in the Moment 1 image, indicating that the whole gas clump is rotating. In the lower panel of Figure 5, we show the Moment 0 contours and Moment 1 images of CH$_{3}$OH (5$_{2,3}$--4$_{1,3}$) emission obtained from combining data from the compact and extended arrays. Under higher spatial resolution, the gas cores in the upper panel fragments into many sub-cores. As depicted by the white dashed line, the sub-cores align in a northwest-to-southeast filamentary structure with a length of $\sim0.5$ pc. While the other sub-cores in the dashed ellipses are located by the side of the filamentary structure. The two solid white lines with arrowheads show the directions of possible jets identified from Spitzer/IRAC image in Figure 1. The directions of the jets are roughly perpendicular to the elongated filamentary structure. The sub-cores in the blue ellipse have much larger blueshifted velocities than the southeast sub-cores of the filamentary structure. The sub-cores in the red ellipse have redshifted velocities.  As depicted by the red dashed lines with arrowheads, the sub-cores in the ellipses spatially connect with the central core and may be spirally falling onto the filamentary structure.

\begin{figure}
\begin{center}
\begin{tabular}{c}
\includegraphics[scale=0.35,angle=0]{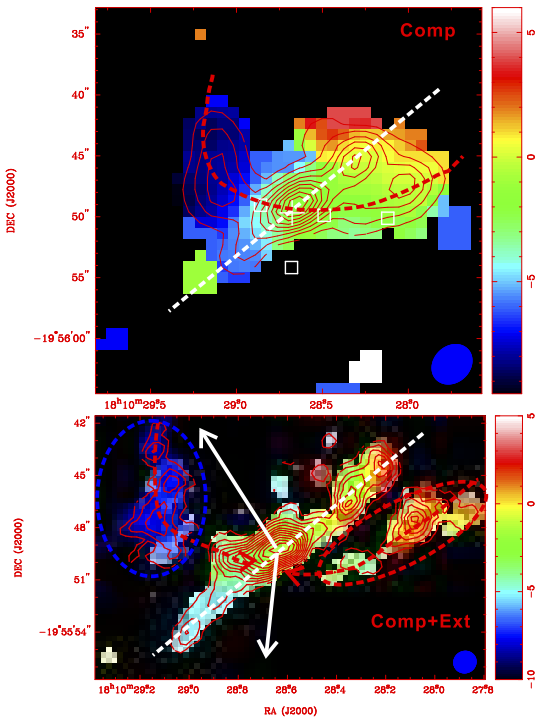}
\end{tabular}
\end{center}
\caption{The contours in both upper and lower panels represent the integrated intensity of CH$_{3}$OH (5$_{2,3}$--4$_{1,3}$). The background are the Moment 1 images of CH$_{3}$OH (5$_{2,3}$--4$_{1,3}$). The data used in the upper panel is from the compact configuration. While in the lower panel, the data is from combining both configurations. The long dashed lines along the NW-SE direction in both panels depicts the rotating plane of the elongated filamentary structure. The solid arrows in the lower panel show the direction of the jets in Figure 1. The other two dashed curves represent the structure of the spiral infalling gas. }
\end{figure}

Is the filamentary structure revealed by CH$_{3}$OH (5$_{2,3}$--4$_{1,3}$) emission really a filament or an edge-on disk-like flatten structure? To answer this question, we need to get the information of its thickness. As shown in Figure 6, we detected seven transitions of CH$_{3}$CN (15-14), which enables us to derive the kinetic temperature and volume density of the gas from LVG modeling. In Figure 7, we presents the integrated intensity map of the summed emission of all the seven CH$_{3}$CN (15-14) transitions. We can see that the CH$_{3}$CN (15-14) emission is mainly associated with the central dust core, indicating the central core is much denser and warmer than the off-centre cores.

We applied RADEX to modeling the CH$_{3}$CN (15-14) emission. In the first running, we explore the parameter space as: H$_{2}$ volume densities n$_{H_{2}}$ of [10$^{6}$,5$\times10^{7}$] cm$^{-3}$, kinetic temperatures T$_{k}$ of [80,200] K and CH$_{3}$CN column densities of [5$\times$10$^{14}$,5$\times10^{17}$] cm$^{-2}$. The we calculated the $\chi^{2}=\sum_{i}\left(\frac{T_{mod}^{i}-T_{obs}^{i}}{\sigma^{i}}\right)^{2}$ , where $T_{mod}^{i},T_{obs}^{i},\sigma^{i}$ of each transition are the modeled brightness temperature, observed brightness temperature and rms of the brightness temperature from gaussian fits. The column density of the CH$_{3}$CN in the best fit is $2\times10^{15}$ cm$^{-2}$, which is consistent with the value obtained from the CH$_{3}^{13}$CN \citep{belt11}. The kinetic temperature and volume density of the best fit are around 160 K and 3$\times10^{6}$ cm$^{-3}$. The we fixed the column density of the CH$_{3}$CN to be $2\times10^{15}$ cm$^{-2}$ and explored the parameter space of n$_{H_{2}}$ and T$_{k}$ in [10$^{6}$,6$\times10^{6}$] cm$^{-3}$ and [140,180] K, respectively. Figure 8 shows the distribution of the normalized $\frac{\chi^{2}}{\chi^{2}_{min}}$, where $\chi^{2}_{min}$ is the smallest $\chi^{2}$ in the running. The best fit gives the n$_{H_{2}}$ of 3$\times10^{6}$ cm$^{-3}$ and T$_{k}$ of 161 K. If we assume the dust temperature of the central core is the same as the kinetic temperature 161 K, then we get the mass and surface density N$_{H_{2}}$ of the central core as 207 M$_{\sun}$ and 3.7$\times10^{24}$ cm$^{-2}$. The thickness the the central core is calculated as $H=\frac{N_{H_{2}}}{n_{H_{2}}}$. We find that the thickness of the central core is $\sim$0.4 pc, which is comparable to the length of the filamentary structure revealed by the CH$_{3}$OH (5$_{2,3}$--4$_{1,3}$). This indicates that the filamentary structure is not a real filament but an edge-on flatten disk-like structure.

\begin{figure}
\begin{center}
\begin{tabular}{c}
\includegraphics[scale=0.25,angle=90]{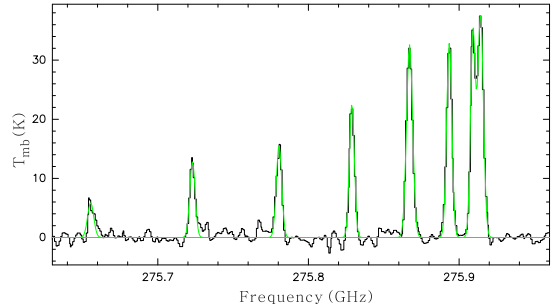}
\end{tabular}
\end{center}
\caption{The beam-averaged spectra of CH$_{3}$CN (15-14) at the position of the emission peak. The green solid lines are the gaussian fits.}
\end{figure}

\begin{figure}
\begin{center}
\begin{tabular}{c}
\includegraphics[scale=0.35,angle=-90]{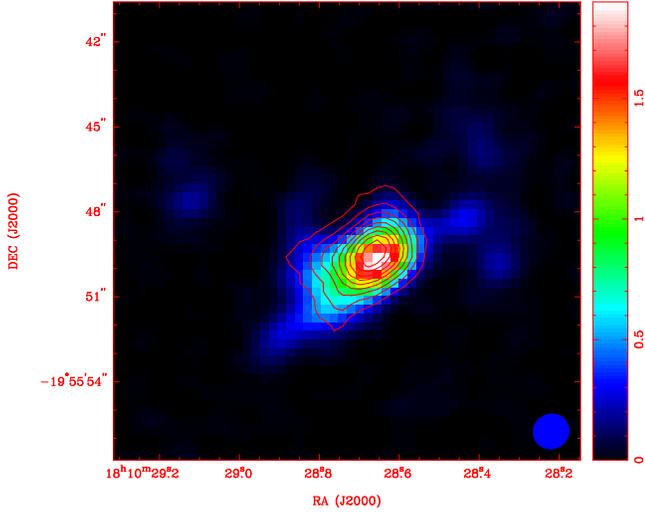}
\end{tabular}
\end{center}
\caption{The integrated intensity of CH$_{3}$CN (15-14). The contours are from 10\% to 90\% in steps of 10\% of the peak value.}
\end{figure}

\begin{figure}
\begin{center}
\begin{tabular}{c}
\includegraphics[scale=0.3,angle=90]{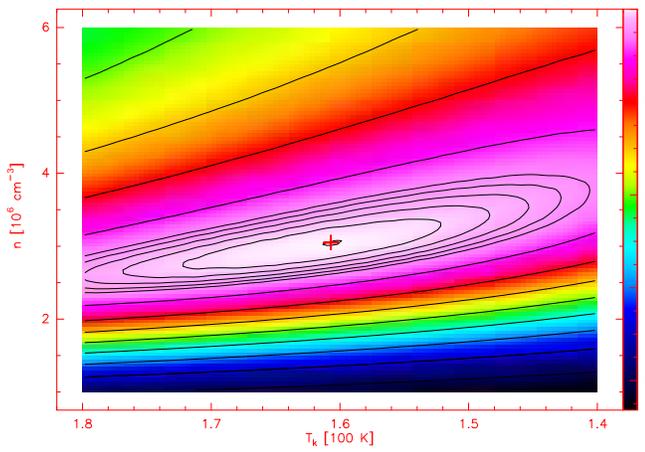}
\end{tabular}
\end{center}
\caption{The volume density (n) and kinetic temperature (T$_{k}$) diagram for CH$_{3}$CN (15-14) from Radex LVG model. The column density of CH$_{3}$CN is 2$\times10^{15}$ cm$^{-2}$. The results of the best fit is shown as a "cross" in the diagram. The corresponding kinetic temperature and the volume density are 161 K
and 3$\times10^{6}$ cm$^{-3}$, respectively. }
\end{figure}

\subsection{Overall collapse}
As shown in the upper panel of Figure 9, the HCN (3--2) spectrum observed by CSO shows blue profile at the centre of the G10.6-0.4 region. Such
a blue asymmetric line profile where the blue emission peak is
at a higher intensity than the red one is a collapse signature \citep{zhou93}. As shown in the lower panel of Figure 9, besides the central position, the HCN (3--2) lines also show blue profile in a much larger region ($\gtrsim20\arcsec$ in radius), indicating that the whole cloud is in overall collapse.

\begin{figure}
  \begin{center}
   \begin{tabular}{c} \includegraphics[width=3in]{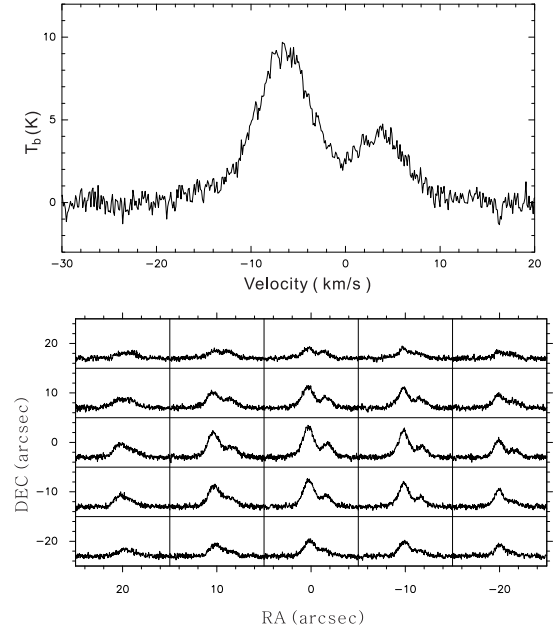}
   \end{tabular}
 \end{center}
 \caption{CSO observations of HCN (3--2)\citep{wu03}. The bottom panel shows the mapping results. The upper panel exhibits the central spectrum.}
 \end{figure}

The redshifted absorption in the HCN (3--2) lines observed by the SMA indicates gas infall in clump scale \citep{welch87}. Interferometers can resolve out the extended emission and create kind of artifacts that may appear like absorption features, especially close to the line centre. However, the absorption dips in the HCN (3--2) line are redshifted by several km~s$^{-1}$ from the systemic velocity, which should not be caused by the missing flux. Especially, the HCN (3--2) line towards core D, F and G shows clearly inverse P-cygni profiles with the absorption dips $\sim$5 K ($\sim 5\sigma$) below the continuum, indicating gas infall towards the bright continuum background \citep{welch87}. In Figure 3, we also noticed that the absorption of HCN (3--2) is associated with the continuum emission, indicating the absorption feature of HCN (3--2) lines are real rather than artificial. From the last panel of Figure 2, one can find that the absorption dip of HCN (3--2) line averaged over the whole clump is at $\sim$1 km~s$^{-1}$. The absorption dip of HCN (3--2) line towards the central core "D" is also at $\sim$1 km~s$^{-1}$. In contrast to "D", the absorption dips of HCN (3--2) lines are slightly blueshifted towards the eastern cores and redshifted towards the western cores, indicating that the infalling gas is from a common cold and slowly rotating envelope. In other words, the clump is in overall collapse. The gas infall is dominated by the whole clump rather than by individual cores.

To investigate the properties of the infalling gas, the profiles of the HCN (3--2) line were modeled with an enhanced version of the "two-layer" method \citep{mye96,fra02}. In that model, a continuum source is located in between the two layers. Each layer has peak optical depth $\tau_{0}$, velocity dispersion $\sigma$,
and relative speed $V_{rel}$ between the continuum source and the gas in the two layers. The gas is approaching the continuum source if $V_{rel}$ is positive. The line brightness temperature
at velocity V is \citep{fra02}:
\begin{eqnarray*}
\Delta T_{B}=(J_{f}-J_{cr})[1-exp(-\tau_{f})]+(1-\Phi)(J_{r}-J_{b})\\
\times[1-exp(-\tau_{r}-\tau_{f})]
\end{eqnarray*}
where
\begin{equation*}
J_{cr}=\Phi J_{c}+(1-\Phi)J_{r}
\end{equation*}
and
\begin{equation*}
\tau_{f}=\tau_{0}exp[\frac{-(V-V_{rel}-V_{cont})^{2}}{2\sigma^{2}}]
\end{equation*}
\begin{equation*}
\tau_{r}=\tau_{0}exp[\frac{-(V+V_{rel}-V_{cont})^{2}}{2\sigma^{2}}]
\end{equation*}
$J_{c}$, $J_{f}$, $J_{r}$, $J_{b}$ are the Planck temperatures of the continuum source, the "front" layer, the "rear" layer
and the cosmic background radiation, respectively. $J=\frac{h\nu}{k}\frac{1}{exp (T_{0}/T)-1}$ is related to the blackbody temperature T at frequency $\nu$, where h is Planck's constant, and k is Boltzmann's constant \citep{mye96}. $\Phi$ and $V_{cont}$ are the filling factors and the systemic velocities of the continuum sources, respectively. The blueshifted emission of HCN (3--2) at core "F" and "G" and redshifted emission of HCN (3--2) at core "B" may be contaminated by outflow emission. Thus we only use the -6 to 5 km~s$^{-1}$ part of the lines in the fitting. We used a nonlinear least-squares method (the Levenberg-Marquardt algorithm coded within IDL) to obtain the best-fit
models for the observational lines. The best fits are displayed in Figure 2 as blue curves and the parameters of the best fits are summarized in Table 2.

From the seventh column of Table 1 and the seventh column of Table 2, we noticed that the systemic velocities of each core obtained from gaussian fits to CH$_{3}$OH (5$_{2,3}$--4$_{1,3}$) lines are slightly different from that obtained from modeling to HCN (3--2) lines. This difference may be due to two factors. Firstly, since CH$_{3}$OH (5$_{2,3}$--4$_{1,3}$) and HCN (3--2) have different critical densities and excitation temperatures, these two molecular lines may trace different layers of the gas. While different layers of the gas may have various velocities. Secondly, the CH$_{3}$OH (5$_{2,3}$--4$_{1,3}$) lines at some positions (e.g. "A" and "E") may be optically thick and deviate from gaussian shape. However, roughly speaking, in contrast with the central core ("D"),  the western cores have redshifted velocities and the eastern cores have blueshifted velocities, indicating the off-centre cores are rotating around the central core.

The last column of Table 2 presents the relative velocities of the cores to the infalling gas. Besides core "E", the relative velocities of the other cores are all much larger than the thermal velocity, indicating the infall motion of the gas is supersonic. For free-fall case, the infall radius can be estimated as:
\begin{equation}
R_{in}=\frac{2GM}{V_{in}^{2}}
\end{equation}
For the central core, the infall velocity is 2.8 km~s$^{-1}$ at a radius of $\sim$0.23 pc. From NH$_{3}$ (3,3) observations, \cite{soll05} detected a free-fall collapse speed of 6.5 km~s$^{-1}$ at a radius of 0.03 pc. This indicates that the inside of the clump collapses faster than outside. The kinematic mass infall rate can be calculated using dM/dt=$4{\pi}n\mu_{G}m_{H_{2}}R_{in}^{2}V_{in}$, where V$_{in}$ is the infall velocity,  $\mu_{G}=1.36$ and $m_{H_{2}}$ are the infall velocity, the mean molecular weight, and the H$_{2}$ mass, respectively. Taking V$_{in}$=6.5 km~s$^{-1}$, R$_{in}$=0.03 pc and n=3$\times10^{6}$ cm$^{-3}$, the kinematic mass infall rate is 1.5$\times10^{-2}$ M$_{\sun}$~yr$^{-1}$.

The "competitive accretion" model predicts the accretion rates in the virialized cluster are primarily determined by the Bondi-Hoyle type accretion rates \citep{bonn02,bonn06}. However, this argument was often doubted and criticized by people \citep{kru05}. The Bondi-Hoyle type accretion rates are believed too low to significantly increase the protostars' masses due to large velocity dispersion in turbulent cores \citep{kru05}. And the radiative feedback from the massive protostars may also prevent Bondi-Hoyle accretions in the cluster \citep{edg04}. Does the mass accretion in G10.6-0.4 favor the Bondi-Hoyle type accretion? The Bondi-Hoyle accretion rate of the central core ("D") is calculated as \citep{kru05}:
\begin{equation}
\dot{M}_{BH}\approx 4\pi \rho \frac{(GM_{*})^{2}}{(\sqrt{3}\sigma)^{3}}
\end{equation}
where G is the Gravitational constant, $\rho$ and $M_{*}$ are the mean density and the stellar mass, respectively. The one dimensional velocity dispersion $\sigma=\sqrt{\sigma_{NT}^{2}+\sigma_{Therm}^{2}}$. For the central core ("D"), the one dimensional thermal velocity dispersion $\sigma_{Therm}=\sqrt{\frac{kT_{k}}{m_{H_{2}}\mu}}$=0.69 km~s$^{-1}$, where $\mu=1.36$ and T$_{k}$=161 K. The one dimensional non-thermal velocity dispersion is estimated from the CH$_{3}$OH (5$_{2,3}$-4$_{1,3}$) line through $\sigma_{NT}=\sqrt{\sigma_{CH_{3}OH}^{2}-\frac{kT_{k}}{m_{CH_{3}OH}}}$=2.17 km~s$^{-1}$. Thus the one dimensional velocity dispersion of the central core is 2.28 km~s$^{-1}$. By inferring the rate of Lyman continuum photon production, \cite{soll05} suggest the central core contains four O4 V stars with a total mass of 175 M$_{\sun}$. Assuming $M_{*}$=175 M$_{\sun}$, the Bondi-Hoyle accretion rate is 2.4$\times10^{-2}$ M$_{\sun}$~yr$^{-1}$, which is similar to the kinematic mass infall rate. However, \cite{kru06} have pointed out that the Bondi-Hoyle formula may underestimate the mean accretion rate in supersonically turbulent gas. But the rate of Bondi-Hoyle accretion onto stars larger than $\sim$10 M$_{\sun}$ is likely to be substantially reduced by radiation pressure\citep{edg04}. Thus our calculation of Bondi-Hoyle accretion rate should be zeroth-order estimation and needs to be greatly modified by considering other factors like vorticity, radiation pressure and magnetic fields\citep{kru06}. However, our analysis indicates that the accretion in the central region is dominated by stellar potential and the central massive core has accumulated enough mass to maintain a large accretion rate even though the velocity dispersion is very larger and supersonic. The ignition of the nuclear reaction can be delayed with such high accretion rate until the star grows massive enough\citep{hos08,hos09}. But the presence of UC H{\sc ii} region indicates that the protostars in the central region of G10.6-0.4 have accumulated large enough mass to produce ionizing radiation. In addition, the central massive protostars seem to continue growing mass by accreting ionized gas passing through the ionization boundary of the ultracompact H{\sc ii} region \citep{keto02}.

\subsection{Competitive accretion in G10.6-0.4?}
In "monolithic collapse and disk accretion", the cores in the clumps collapse individually to form stars. However, in G10.6-0.4 region, the cloud/clump seems to be in overall collapse. The central dense core (core "D") has much larger mass than the off-centre cores. In free-fall case, the accretion domain ($R_{in}\sim$47440 AU) of the central core is much larger than the clump size ($\sim$10000 AU), indicating that the central dense core dominates the gravity of the clump and thus can competitively gather gas mass with the other off-centre protostars from the gas reservoir of the clump. We also noticed the gas infall in this region is dominated by the central protostars in Bondi-Hoyle accretion mode. The difference between the "competitive accretion" model and the "monolithic collapse and disk accretion" lies primarily in how and when the mass is gathered to form the massive star \citep{bonn08}. Does it occur in the pre-stellar stage due to turbulence, or does it occur subsequently due to accretion in a common potential \citep{bonn08}. G10.6-0.4 is a much evolved protocluster. But the protostars and their gas reservoirs (dense cores) in G10.6-0.4 are still accumulating masses from the gas reservoir of the clump, indicating that the cores in G10.6-0.4 are most likely formed due to accretion rather than pre-existing. \cite{liu12} detected several 5 pc scale filaments connected with the central region of G10.6-0.4. They argued that the filaments may facilitate mass accretion onto the central cluster-forming region. The overall collapse of the clump and the large scale filaments indicate that the mutual potential of the protocluster likely funnels gas from the natal cloud down to the centre. This picture favors the "competitive accretion" model.

The standard deviations of core velocities measured from CH$_{3}$OH (5$_{2,3}$--4$_{1,3}$) lines and HCN (3--2) lines are $\sim$2.5 and $\sim$0.8 km~s$^{-1}$, both of which are larger than the thermal velocity dispersion of the clump ($\sim$0.5 km~s$^{-1}$ assuming T$_{k}$=87 K). The velocity differences between most of the cores and the cloud/clump (-3 km~s$^{-1}$) are also higher than the thermal velocity dispersion. From the last column of Table 1, we noticed the non-thermal velocity dispersion of the cores are much larger than the thermal velocity dispersion. In "monolithic collapse and disk accretion", the cores in a clump do not interact with each other and collapse individually, leading to low sonic velocity dispersion of the gas and small velocity differences between cores and their envelopes\citep{Kru09}. While the velocity dispersion of the gas increases with mass and can reach as high as several km~s$^{-1}$ in the "competitive accretion" model\citep{bonn06}. The "competitive accretion" model also predicts high relative gas velocities in a stellar-dominated potential.

At the present time, we can not make unambivalent conclusion that the protocluster in G10.6-0.4 is formed through competitive accretion due to uncertainties in estimating the masses and mass infall rates. However, we clearly revealed the mass segregation in this region and the cores seem to still accumulate masses due to a common potential. Further higher sensitivity and resolution observations especially from the ALMA can give a more thorough understanding of the situations in this region especially of the interaction among the individual cores.

\section{Summary}
We present the SMA observations both in the 1.1 mm continuum and molecular line emission toward the high-mass star forming complex in G10.6-0.4.
The main results of this study are as follows.

1. The 1.1 mm dust emission fragments into seven dense cores. The central one has the largest mass of $\sim$207 M$_{\sun}$, while the off-centre ones have much smaller masses.

2. The gas emission reveals by HCN (3--2) and CH$_{3}$OH (5$_{2,3}$--4$_{1,3}$) lines also fragments into several dense cores. The dense cores revealed by the CH$_{3}$OH (5$_{2,3}$--4$_{1,3}$) emission align along a filamentary structure with a length of $\sim$0.5 pc. The kinetic temperature and volume density of the central core inferred from LVG modeling to CH$_{3}$CN (15-14) lines are $\sim$161 K and $3\times10^{6}$ cm$^{-3}$. Thus the thickness of the filamentary structure in CH$_{3}$OH (5$_{2,3}$--4$_{1,3}$) emission is $\sim$0.4 pc, which is comparable to its length, indicating the filamentary structure is not a real filament but an edge-on disk-like structure. There exists a large velocity gradient from west to east, indicating the whole clump is in rotating.

3. The HCN (3--2) lines towards all the dust cores show redshifted absorption dips (inverse P-Cygni profile), indicating gas infall in this region. The positions of the absorption dips at different cores are nearly at the same velocity ($\sim$1 km~s$^{-1}$), indicating that the collapse is takeing place in clump scale rather than core scale. In other words, the gas is accreted due to the common potential of the proto-cluster rather than due to individual cores. The single-dish observations of HCN (3--2) also support large scale collapse in cloud scale. We also found that the Bondi-Hoyle accretion can explain the gas accretion in the forming cluster of G10.6-0.4.

4. The non-thermal motion in this region is supersonic. The standard deviation of core velocities is also larger than the thermal velocity dispersion. The velocity differences between the cores and the cloud/clump are also larger than the thermal velocity dispersion. Through modeling the HCN (3--2) lines, we found the relative velocities between the dust continuum sources and the infalling gas are much larger than thermal velocity dispersion, indicating the infall is supersonic.

5. The observed picture in G10.6-0.4 region seems to favor the "competitive accretion model", which needs to be tested by future observations.

\section*{Acknowledgments}

We are grateful to the SMA staff. This work was funded by China Ministry of Science and Technology under State Key Development Program for Basic Research 2012CB821800 and the NSFC grant 11073003. We also thank the referee for the useful comments.

\end{document}